\documentclass[10pt, conference, compsocconf]{IEEEtran}
\ifCLASSINFOpdf
  \usepackage[pdftex]{graphicx}
\else
\fi
\usepackage{algorithmic}
\begin{document}
%
\title{An Incremental Local-First Community Detection Method for Dynamic Graphs}


\author{\IEEEauthorblockN{Hiroki Kanezashi}
\IEEEauthorblockA{Tokyo Institute of Technology \\
IBM T.J. Watson Research Center \\
2-12-1 Oo-okayama Meguro Tokyo, JAPAN \\
kanezashi.h.aa@m.titech.ac.jp}
\and
\IEEEauthorblockN{Toyotaro Suzumura}
\IEEEauthorblockA{IBM T.J. Watson Research Center \\
Barcelona Supercomputing Center, The University of Tokyo \\
New York, USA \\
suzumura@acm.org}
}


%


\maketitle

\begin{abstract}
Community detections for large-scale real world networks have been more popular in social analytics. In particular, dynamically growing network analyses become important to find long-term trends and detect anomalies.
In order to analyze such networks, we need to obtain many snapshots and apply same analytic methods to them. However, it is inefficient to extract communities from these whole newly generated networks with little differences every time, and then it is impossible to follow the network growths in the real time.
We proposed an incremental community detection algorithm for high-volume graph streams. It is based on the top of a well-known batch-oriented algorithm named DEMON \cite{coscia2012demon}.
We also evaluated performance and precisions of our proposed incremental algorithm with real-world big networks with up to 410,236 vertices and 2,439,437 edges and computed in less than one second to detect communities in an incremental fashion - which achieves up to 107 times faster than the original algorithm without sacrificing accuracies.
\end{abstract}


%
\IEEEpeerreviewmaketitle

\section{Introduction}
Community detections for large-scale networks have been important research problems in graph computing, and real-time analyses of complex social networks are also important to discover trends.
Most of real-world networks like user-relationships on social networking services are time-evolving and dynamic networks.
Consequently, in order to detect and extract communities from large-scale and time-evolving dynamic social network in nearly real-time fashion, it is inefficient to compute them from the beginning of the entire network every time, even though there is little change. To overcome the problem, several approaches have been proposed for community detection algorithms working in an incremental fashion, which is often called incremental community detection.

We propose an incremental community detection algorithm  based on one of well-known and scalable community detection algorithms named DEMON \cite{coscia2012demon} for incrementally growing networks.
We implemented the incremental algorithm in C++, and conducted performance evaluations with real-world networks representing relations of people. Then we showed that our method took only less than a few seconds to compute incrementally updated networks while original DEMON needed to execute community detections for a whole network in a batch fashion - which brought tremendous amount of overhead.

Our contributions are following: we proposed an incremental version of DEMON algorithm with DEMON devising core functions, reduced unnecessary community comparison procedures by defining a community mapping table, and proved the effect of these proposed optimizations using real-world networks.

The next section will describe related works about incremental community detection. Section 3 will introduce the original DEMON algorithm and its three core functions our method based on. Our proposed incremental DEMON algorithm will be introduced in Section 4, and the result of performance evaluation of our method will be described in Section 5. Finally, we will summarize our research and future works in Section 6.

\section{Related Work}
Zhenyu and Ming propose a scalable community detection method for tag assignments stream clustering (TASC)\cite{wu2014incremental}.
The method requires not only network structures, but needs tag set and user data for all network vertices. They used three types of data sets with about 10,000 to 70,000 vertices and 50,000 to 440,000 tag assignments, and no initial edges were specified. Their performance experiment showed that it took about 10 ms to 30 ms with one assignment.

Sheng et. al. focused an incremental label propagation algorithm and added the time sequences as well as labels to vertices \cite{pang2009realtime}.
The average computation complexity was the same as the total number of local edges connected to the local vertices $O(m)$ (m: the number of edges), and the worst case of the overall time was $O(m + n)$ (n: the number of vertices). 
They showed that the average iteration time was 4.411 seconds with on-line game players networks. In this algorithm, additional tag data were necessary and the final results of network division might change with the number of iterations. Our proposed method considers only network structures, and output results are stable.

Jingyong and Hongsheng proposed an incremental algorithm based on growing social networks considering that most communities tended to evolve gradually over time\cite{li2012cdbia}.
They used computer-generated data sets and internet peer-to-peer network data with 10,876 vertices and 39,994 edges for their experiments.
Compared with RI (based on Radicci method) and IC (finding nodes that are connected with increments: almost same as this method), the computation time was almost same as the IC method, and better than the other methods especially increasing the total number of vertices.

\section{Background - Overlapping Community Detection}
In this section we will introduce an overlapping community detection algorithm called DEMON, since our proposed incremental algorithm uses this as a base algorithm.

\subsection{DEMON Overview}
DEMON \cite{coscia2012demon} is a scalable algorithm for detecting overlapping communities in complex networks. In this context, a community is a group of nodes densely connected to each other. It outputs overlapping communities, that means some nodes will be in different communities at the same time. We choose this algorithm as baseline of our research with several reasons.

First, DEMON has fine-grained scalability. Because communities will be generated in a bottom-up way and some of the required computation for all the vertices can be independently executed, it would fit with parallel and distribution computation.
We considers it is also suitable for incremental extension.

Second, it is the state-of-the-art community detection algorithm that outperforms other algorithms in terms of accuracy. DEMON algorithm allows overlapped community outputs, that is usual for real-world networks.
Most of large-scale social networks contains many vertices with multiple properties, so DEMON algorithm should output proper communities from these properties. Moreover, The author of \cite{coscia2012demon} showed DEMON provided higher accuracy than other community detection algorithms in that the F-Measure scores based on labels of vertices are the largest. They also described that the community size distribution of DEMON were more balanced than that of Infomap algorithm.

Furthermore, three steps of DEMON algorithm including {\it Ego Minus Ego}, {\it Label Propagation} and {\it Merge} functions are performed for all vertices in an independent manner. Because each vertex is processed independently, some functions could be executed by incremental procedures.

\subsection{Algorithm Details}
DEMON algorithm repeats the following functions including {\it Extracting Ego Minus Ego Networks}, {\it Label Propagation} and {\it Merge} functions for each vertex in the input network.
\begin{enumerate}
\item Extract an {\it Ego Minus Ego} network from a vertex.
\item Group vertices from the extracted {\it Ego Minus Ego} network according to the result of label propagation.
\item Merge generated groups by the overlapping degree.
\end{enumerate}

Next, we will describe details of these three functions DEMON algorithm consists of.

\subsubsection{Extracting Ego Minus Ego Networks}
In DEMON algorithm, the first step extracts an Ego Network for the chosen vertex. Ego Network is a sub-graph of original input network consists of the specified vertex (Ego), vertices neighboring the Ego vertex and edges connecting these vertices.
Obviously a single node connecting the entire sub-graph are connected directly and affect the similarities of neighboring vertices, even if they are not in the same community. For this reason, the ego vertex is removed from its own ego network in this function. {\it Ego Minus Ego} network is a sub-graph removed the Ego vertex from the Ego Network.

\subsubsection{Label Propagation}
After the {\it Ego Minus Ego} graph is extracted from the specified vertex, the next step is to compute communities subsets of this network by {\it Label Propagation}. The label propagation method is based on \cite{raghavan2007near}, updating vertices labels by other label frequency of neighbor vertices.

This function repeats the following procedures to find small communities with the graph structure.
\begin{enumerate}
\item Set numerical labels to all vertices. Each label must be different from those of other vertices.
\item Set time stamp $t = 1$.
\item Choose a vertex from the network randomly and put labels gather from neighboring vertices.
\item Find the majority of gathered labels and set it as the new label of this vertex.
\item After processing all vertices, if all labels are larger number than those of neighbors or time stamp reaches to the upper limit, finish this label propagation algorithm.
\item Otherwise, increment the time stamp ($t = t + 1$) and process vertices again.
\end{enumerate}

\subsubsection{Merge}
{\it Label Propagation} function generates many small communities and they are restricted to be sub-graphs of {\it Ego Minus Ego} networks. In order to find the larger and global communities in the given whole graph, they will be merged in {\it Merge} function.

In {\it Merge} function, generated communities by {\it Label Propagation} will be merged. Parameter $\epsilon$ is a given threshold indicating overlaps between communities.
They will be merged if and only if at most the $\epsilon$ of the smaller one is not included in the bigger one. When $\epsilon = 0$, two communities will be merged only if one of them is a proper subset of the other, and when $\epsilon = 1$, they will be merged together even these communities do not share any nodes. In general, two overlapped community will be more likely merged if the value of $\epsilon$ is large.

\begin{figure}[htb]
    \centering
    \includegraphics[width=0.8\hsize]{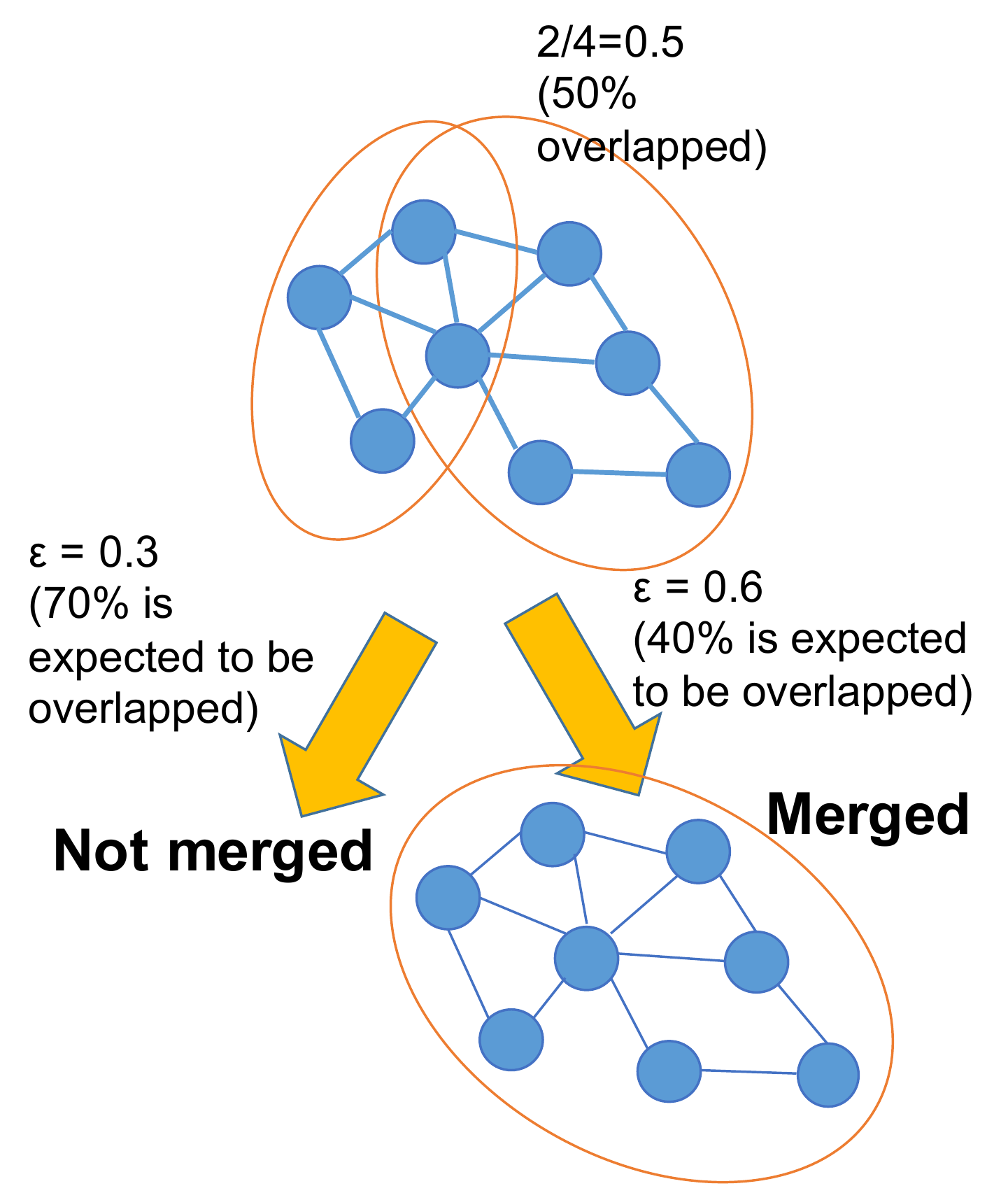}
    \caption{An Example of {\it Merge} Function.}
    \label{fig:merge}
\end{figure}

In this example described in Figure \ref{fig:merge}, two communities are overlapping each other. The left one has 4 nodes, the right one has 6 nodes and 2 vertices are overlapped. In this situation, they are 50\% ($\frac{2}{4}$) overlapped.
If $\epsilon = 0.6$, at least 40\% vertices of the smaller community must be overlapped to be merged, then these communities will be merged as a community with 8 vertices. On the other hand, if $\epsilon = 0.3$, at least 70\% vertices must be overlapped for merge, then they will not be merged.

The {\it Merge} function also randomly chooses an existing community to be chosen next by shuffling orders of communities to prevent choosing the same communities in every iterations, which causes unequal output of communities.

\subsection{Performance Characteristics of DEMON algorithm}
Coscia et. al. \cite{coscia2012demon} evaluated their original DEMON algorithm with real-world networks in Java implementation.
In these performance evaluations, they used Amazon data set with 410,236 vertices and 2,439,437 edges.
They mentioned the core of DEMON algorithm ({\it Ego Minus Ego} and {\it Label Propagation}) took less than a minute, while Merge function with increasing thresholds elapsed one minute to one hour. 

In Congress and IMDb data sets, the original DEMON program detected the better community in F-Measure than other referred algorithms (HLC \cite{ahn2010link}, Infomap \cite{rosvall2008maps}, Modularity \cite{clauset2004finding} and Walktrap \cite{pons2005computing}). However, they did not evaluate more detailed DEMON's performance comparing with those algorithms or network data.

\section{Proposed Method - Incremental Community Detection}
We propose an incremental version of DEMON algorithm named "Incremental DEMON" that incrementally detects communities over time from streaming data.
Incremental DEMON is comprised of three functions: incremental {\it Ego Minus Ego} network extraction and modification, incremental {\it Label Propagation} and optimized {\it Merge} function.

\subsection{Incremental Functions Description}
In the original DEMON algorithm, intermediate networks and communities are generated for all vertices (ego) in the given graph from the first state. However, most of the results of {\it Ego Minus Ego} extractions, {\it Label Propagation} and {\it Merge} procedures from the incrementally updated network are the same as from based networks if there are few differences between network difference like only additional a couple of vertices or edges with the real social network growth in a few seconds.

In this situation, it is time-wasting to repeat same graph processing every time and impossible to catch up a real-time network growths. To overcome this problem, we proposed an incremental version of DEMON by introducing incremental algorithms for these three functions, and enabled it to minimum modification for existing communities. We will show incremental versions of them.

\subsubsection{Incremental Ego Minus Ego Function}
\begin{figure}[htb]
    \centering
    \includegraphics[width=\hsize]{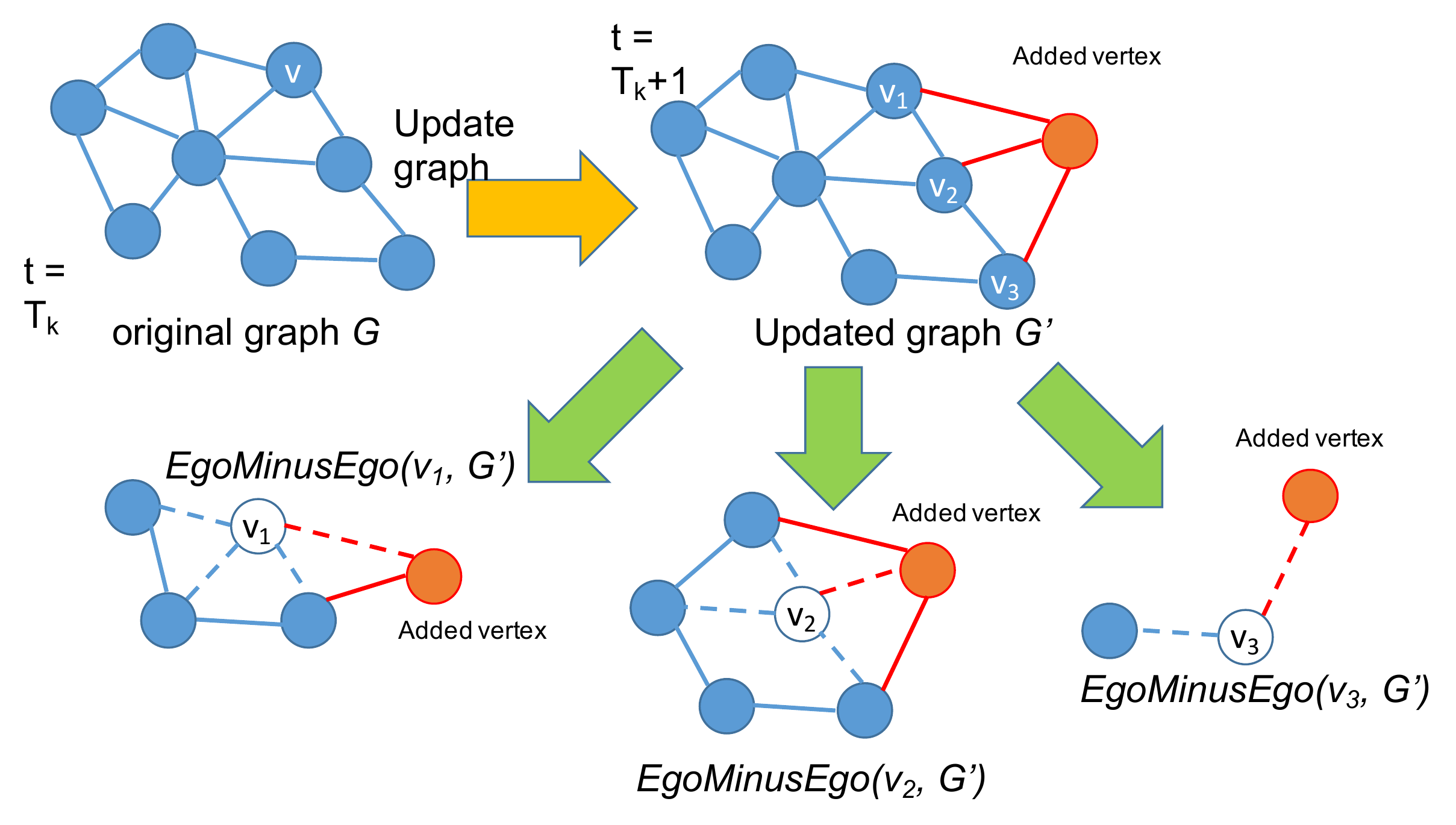}
    \caption{An Example of Incremental {\it Ego Minus Ego} (add single vertex and neighboring edges).}
    \label{fig:eme}
\end{figure}

In the incremental situation, An added vertex or edge will affect only the neighbor vertices, and almost all {\it Ego Minus Ego} networks will not be changed before and after the incremental process.
When a new vertex is added, we need only to re-construct {\it Ego Minus Ego} networks for the neighboring vertices of the newly added vertex like Figure \ref{fig:eme}.
When a vertex is removed, we need only to remove a vertex and edges from {\it Ego Minus Ego} networks for the neighboring vertices of the removed vertex.

The pseudo code of the incremental {\it Ego Minus Ego} function is shown in Algorithm \ref{inc-eme}. When single vertex and connecting edges to neighbors are newly added, The algorithm is repeated for each added edge.
Compared with the original {\it Ego Minus Ego} function coustructing {\it Ego Minus Ego} networks for all vertices (Algorithm \ref{orig-eme}), the incremental function only updates two vertices for origin and destination of an newly added edge.

\begin{figure}[htb]
\label{orig-eme}
\begin{algorithmic}
\REQUIRE $G:(V, E), v \in V $
\ENSURE $G': (V', E')$
\FORALL{$v \in V$}
\STATE{$V' \leftarrow neighbors(v)$}
\FORALL{$v' \in V'$}
\FORALL{$e' \in outer(v')$}
\IF{$target(e') = v$}
\STATE{$E' \leftarrow E' \cup e'$}
\ENDIF
\ENDFOR
\ENDFOR
\ENDFOR
\end{algorithmic}
\caption{Original {\it Ego Minus Ego} algorithm}
\end{figure}

\begin{figure}[htb]
\label{inc-eme}
\begin{algorithmic}
\REQUIRE $G:(V, E), e \in E,$\\ ${\it Ego Minus Ego}:G'=(V', E') \subseteq G$
\ENSURE Updated $G'$
\FORALL{$v \in [source(e), target(e)]$}
\STATE{$V' \leftarrow neighbors(v)$}
\FORALL{$v' \in V'$}
\FORALL{$e' \in outer(v')$}
\IF{$target(e') \in V'$}
\STATE{$E' \leftarrow E' \cup e'$}
\ENDIF
\ENDFOR
\ENDFOR
\ENDFOR
\end{algorithmic}
\caption{Incremental {\it Ego Minus Ego} algorithm (add single edge)}
\end{figure}

\subsubsection{Incremental Label Propagation Function}
The incremental label propagation for each extracted {\it Ego Minus Ego} network will perform independently.
At {\it Label Propagation} algorithm, a frequently appeared vertex is selected or if the frequency is the same, and then it randomly selects a vertex. Repeat the {\it Label Propagation} algorithm only near the added / removed vertex.
Because the idea of label propagation method in original DEMON is based on \cite{raghavan2007near}, which chooses and sets the label with highest frequency in neighbor vertices, it is natural to set the highest flequently appeared labels to those added vertices.

\begin{figure}[htb]
    \centering
    \includegraphics[width=0.7\hsize]{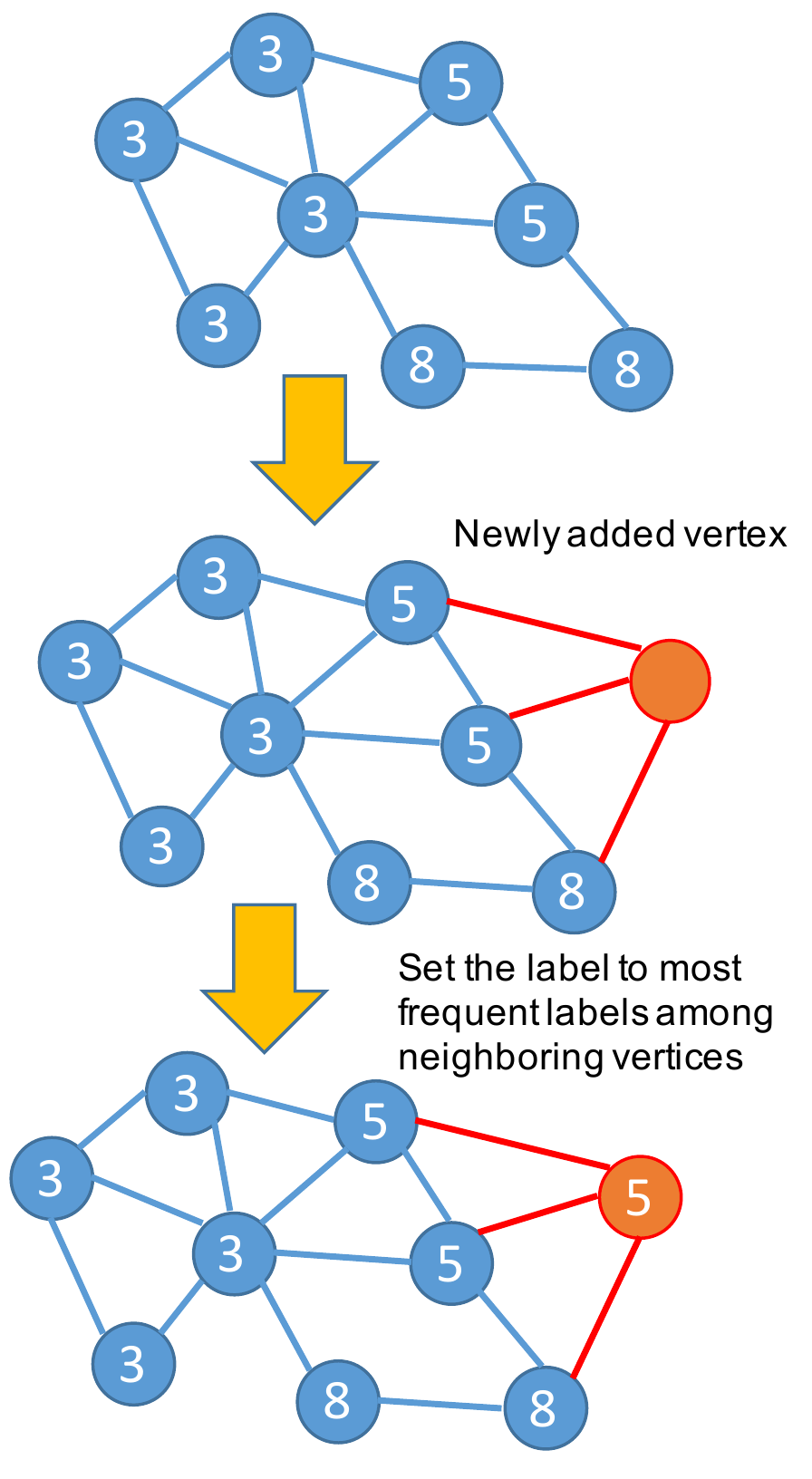}
    \caption{Incremental Label Propagation.}
    \label{fig:lp}
\end{figure}

An example of incremental {\it Label Propagation} procedure with addition a vertex is shown Figure \ref{fig:lp}.

\subsubsection{Merge Function}
DEMON algorithm applies "local-first" approach rather than top-down view. Generated communities from previous {\it Label Propagation} procedure are too small and local. In order to detect more global communities, all overlapping communities should be merged together. Needless to say, the result of incremental {\it Label Propagation} is incrementally changed, so we need to apply a little modification to the original result of {\it Merge} process.

The authors of \cite{coscia2012demon} mentioned, however, the Merge function does not hold the incrementality property, and an alternative simpler algorithm keeps incrementality were needed as their future work.

On the other hand, an alternative Merge function can be defined to combine the results provided by the core of the algorithm, thus preserving its possibility to scale up in a parallel framework. 
We cannot adjust the result of merge function only by adding or removing sub-graphs because the decisions whether some communities should be merged would change and future scenarios might be different.

On the other hand, most communities with no common vertices do not have any relationships with added single vertex, and we can reduce the number of comparisons between an incrementally updated community and other communities.
In order to specify communities which each vertex belongs to, we defined a mapping table between vertex ID and set of community ID.
Elements of the table will be constructed and refereed during the initial {\it Merge} function and the construction cost of this table object can be ignored in the incremental process by reusing it.

We describe optimized merge function to Algorithm \ref{merge-opt}.

\begin{figure}[htb]
\begin{algorithmic}
\REQUIRE $C$ : Total community set,\\ $c$ : Updated community,\\ $T$ : Mapping table,\\ $\epsilon$ : Threshold
\ENSURE Updated $C$ and $T$
\STATE{$Cset \leftarrow \phi$}
\FORALL{$v \in c$}
\STATE{$Cset \leftarrow Cset \cup T[v]$}
\ENDFOR
\STATE{$merged \leftarrow False$}
\FORALL{$c' \in Cset$}
\IF{$c'.size \leq c.size\ and\ c' \subseteq_{\epsilon} c$}
\STATE{$u \leftarrow c' \cup c$}
\STATE{$C - c',\ C - c$}
\STATE{$C \leftarrow C \cup c$}
\STATE{$merged \leftarrow True$}
\STATE{Return}
\ENDIF
\ENDFOR
\IF{$merged = False$}
\STATE{$C = C \cup c$}
\FORALL{$v \in c$}
\STATE{$T[v] \leftarrow T[v] \cup c$}
\ENDFOR
\ENDIF
\end{algorithmic}
\caption{Pseudo Code of Optimized {\it Merge} Algorithm with Community Mapping Tables}
\label{merge-opt}
\end{figure}

In the first for-all loop, this function extracts possible communities to compare. Keys of the mapping table $T$ are vertex IDs and values are sets of community ID which the vertex belongs to ($T[v]$ means the community ID sets with vertex $v$). At the end of this for-all loop, the community set $Cset$ stores the all possible community IDs which might be merged to the given community $c$.

These communities are compared in the second for-all loop one after another. The inside of this loop is the exactly same as the comparison procedure of {\it Merge} function in original DEMON algorithm. If a community should be merged is found, merge these communities and finish this optimized {\it Merge} function. If there are no communities to be merged, register the given community $c$ as the new global community and update this mapping table.

In fact, our optimized method has extra cost for constructions and references the mapping table. However, the number of comparison and merge process were reduced by referring this table.
While original {\it Merge} function compared all-to-all small communities even most communities are never overlapped, our optimized {\it Merge} function compares only overlapping communities, then it eliminates unnecessary merging repetition.

Furthermore, we can reduce the community comparison processes in the incremental DEMON algorithm.
In order to find communities we should actually compare quickly, we defined two types of tables.
The first table represents mapping between vertex ID and community IDs the vertex belongs, and the second one represents the number of vertices and overlapping communities for each community.
These tables will be updated when vertices or edges are incrementally added to or deleted from communities.

\begin{figure*}[htb]
    \centering
    \includegraphics[width=0.9\hsize]{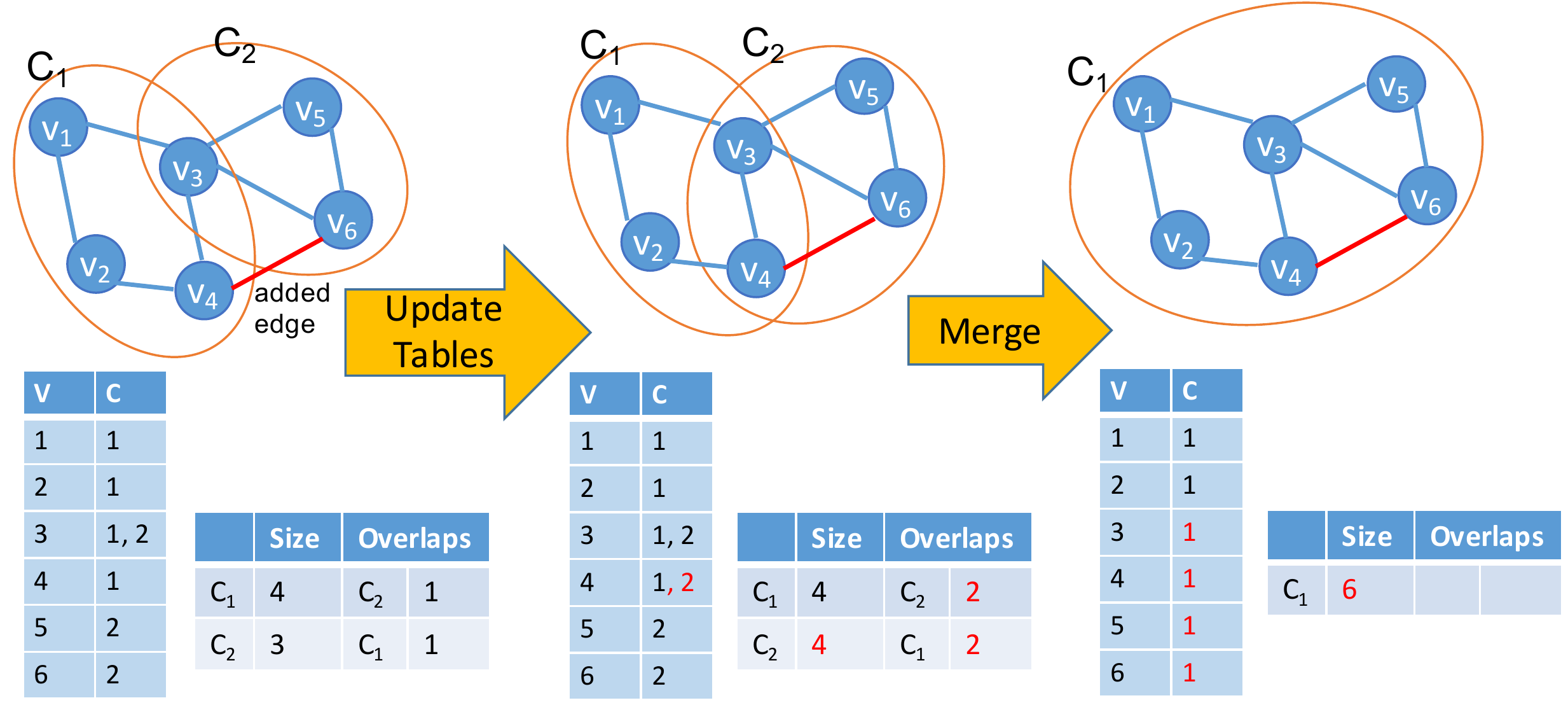}
    \caption{An Example of Incremental Merge Processing with Optimized Data Structures.}
    \label{fig:mergeOpt}
\end{figure*}

We will show an example of the way with how incremental {\it Merge} processing with updates of these tables when an edge is added (Figure \ref{fig:mergeOpt}).

\begin{enumerate}
\item Suppose an edge between $v_4$ and $v_6$ is added, and $v_4$ is about to join community $C_2$ by the result of incremental {\it Label Propagation} function.
\item Update these tables describe member vertices and overlaps according to updated vertices and communities where they belongs.
\item Apply {\it Merge} function only to communities which of statuses are updated.
\item If they are actually merged, update these mapping tables again.
\end{enumerate}

\subsection{Complexity}
Suppose the number of degrees of the updated (added or removed) vertex $v$ is $k_v$ and the average degree of graph is $k$. $n$ and $m$ is the number of total vertices and edges of a given graph. We also define $|C|$ as the number of generated communities.

\subsubsection{Ego Minus Ego Network Extraction}
The {\it Ego Minus Ego} function extracts {\it Ego Minus Ego} networks from neighbor vertices of the update vertex, and then add or remove the neighbor vertices. From the discussion, the computation complexity for an {\it Ego Minus Ego} function is $O(k_v \times k)$.

Considering that {\it Ego Minus Ego} networks must be stored for all vertices, the space complexity is $O(n+m)\times k$. That means it needs large heap memory space to store the number of average degree of a graph, and it is impossible to manage billion-scale networks in this method. We will need a scalable method such as data optimization and leveraging distributed-memory environments for our future work to handle larger-scale data.

\subsubsection{Label Propagation}
The {\it Label Propagation} function puts or updates a label to the newly added vertex. The computation complexity is $O(k_v)$ because it only updates the added vertex label from neighbor vertices.

The space complexity is only $O(n \times k^2)$, because the function stores the member of communities for each {\it Ego Minus Ego} network. The required heap memory space might be much larger than {\it Ego Minus Ego} network extraction.

\subsubsection{Merge Function}
In the original DEMON, the computation and space complexity in the {\it Merge} function is $|C|$ and $|C| \times n$. On the other hand, our optimized {\it Merge} function, the computation and space complexity is up to $|k_v|$ and $|k_v| \times n$ because we have only to compare near communities from neighbor vertices.

\section{Implementation and Evaluation}

\subsection{Implementation}
We implemented original and incremental DEMON algorithms in C++ on top of our own graph database called IBM System G \cite{systemg}. The original DEMON algorithm has been implemented as a Python script and can be downloaded from DEMON homepage \cite{demonweb}, and we ported this script to C++ code.

\subsection{Experimental Data Set}
In order to compare our proposed incremental method and the original one, we obtained the same data sets as the experiment of the authors of \cite{coscia2012demon} did as follows. 

\begin{description}
\item[Congress]\mbox{}\\
    The network of legislative collaborations between US representatives of the House and the Senate during the 111st US congress. It is a relatively dense network, with 536 vertices and 14,198 edges, and average degree is 53.98.
\item[IMDB]\mbox{}\\
    We also used the part of the data set of actors who star in at least two movies during the years from 2001 to 2010, filtering out television shows, video games, and other performances. The number of vertices and edges are 56,542 and 185,247. The average degree is 6.55.
\item[Amazon]\mbox{}\\
    As a larger real network, we used Amazon purchases data. It has frequent co-purchases of products are recorded for the day of May 5th 2003. The number of vertices and edges are 410,236 and 2,439,437, and the average degree is 11.89.
\end{description}

Input files for the experiments have edge list in the CSV format where each line is comprised of one edge from a source vertex to a destination vertex. In order to simulate time-evolving dynamic graph, we have 2 different files - one for baseline graph data and another file for dynamically added edge list. Base graph data to be added will be read at once and additional edge data to be added incrementally.

\subsection{Experimental Environment}
In the original DEMON algorithm, we executed the whole algorithm for dynamic networks. We generated each network input file by newly added edge data one by one, then start DEMON program and measure execution time and obtain communities.
On the other hand, our incremental DEMON algorithm loads a baseline graph data and calculate the initial community only once.
We evaluated performance by executing the original and incremental DEMON algorithm on IBM System G process. The execution environment is 64bit Red Hat Linux, with single core Intel(R) Xeon(R) CPU E5-2660 @ 2.20GHz and 378GB random access memory.

\subsection{Performance Results}
\begin{figure}[htbp]
    \centering
    \includegraphics[width=\hsize]{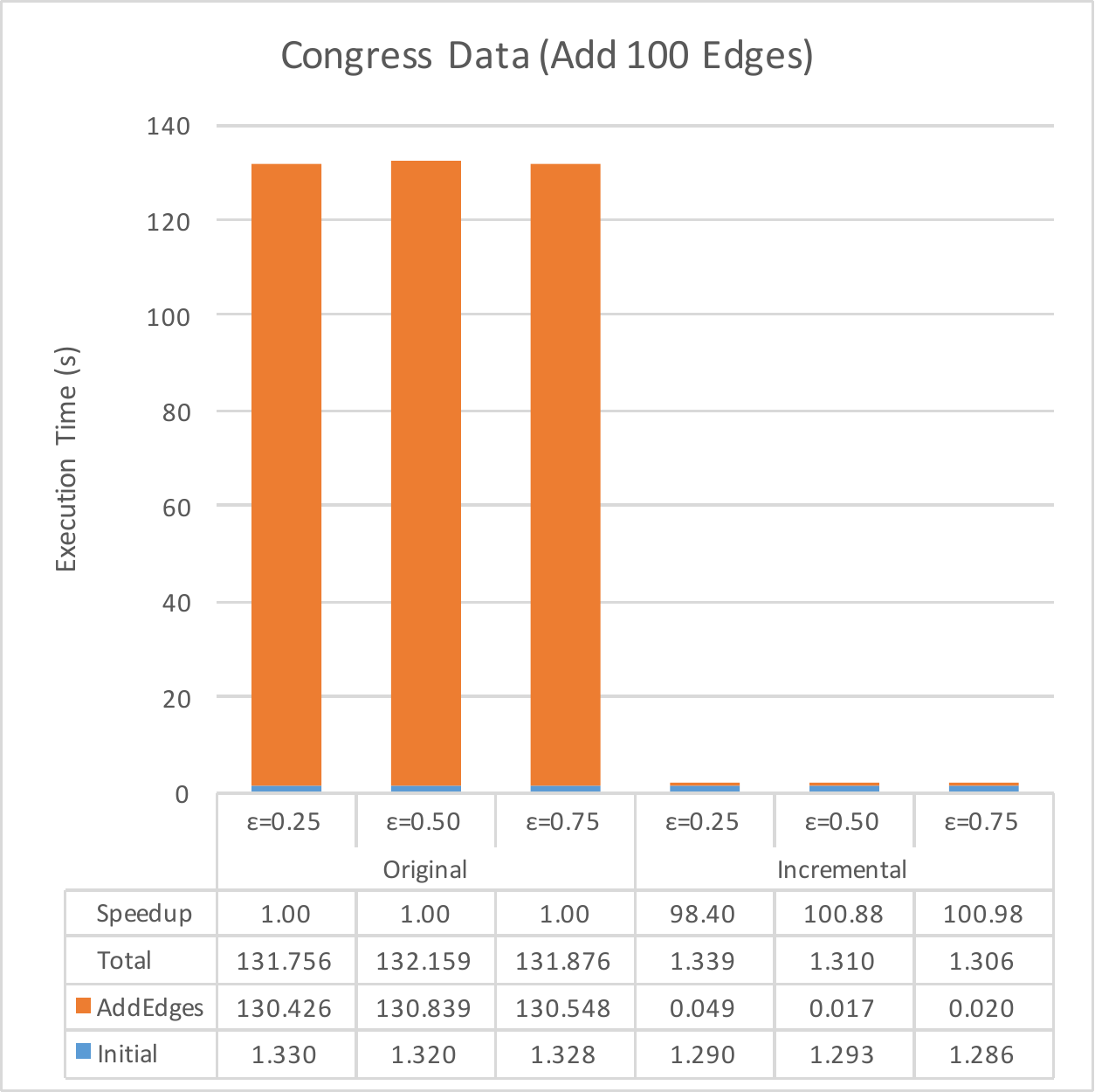}
    \caption{Elapsed time using Congress graph data when new 100 edges are added}
    \label{fig:congress}
\end{figure}

\begin{figure}[htbp]
    \centering
    \includegraphics[width=\hsize]{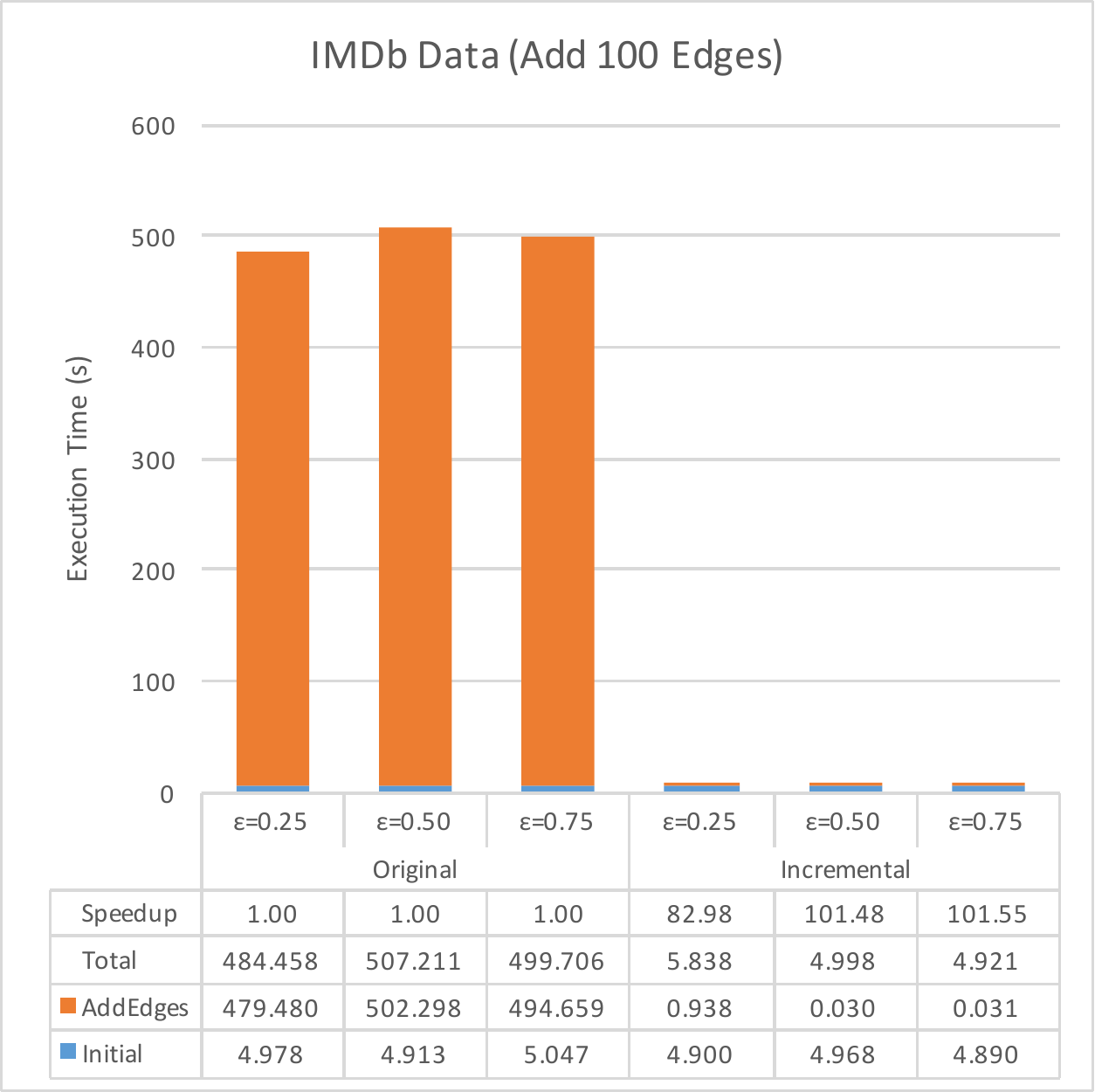}
    \caption{Elapsed time using IMDb graph data with additional 100 edges}
    \label{fig:imdb}
\end{figure}

\begin{figure}[htbp]
    \centering
    \includegraphics[width=\hsize]{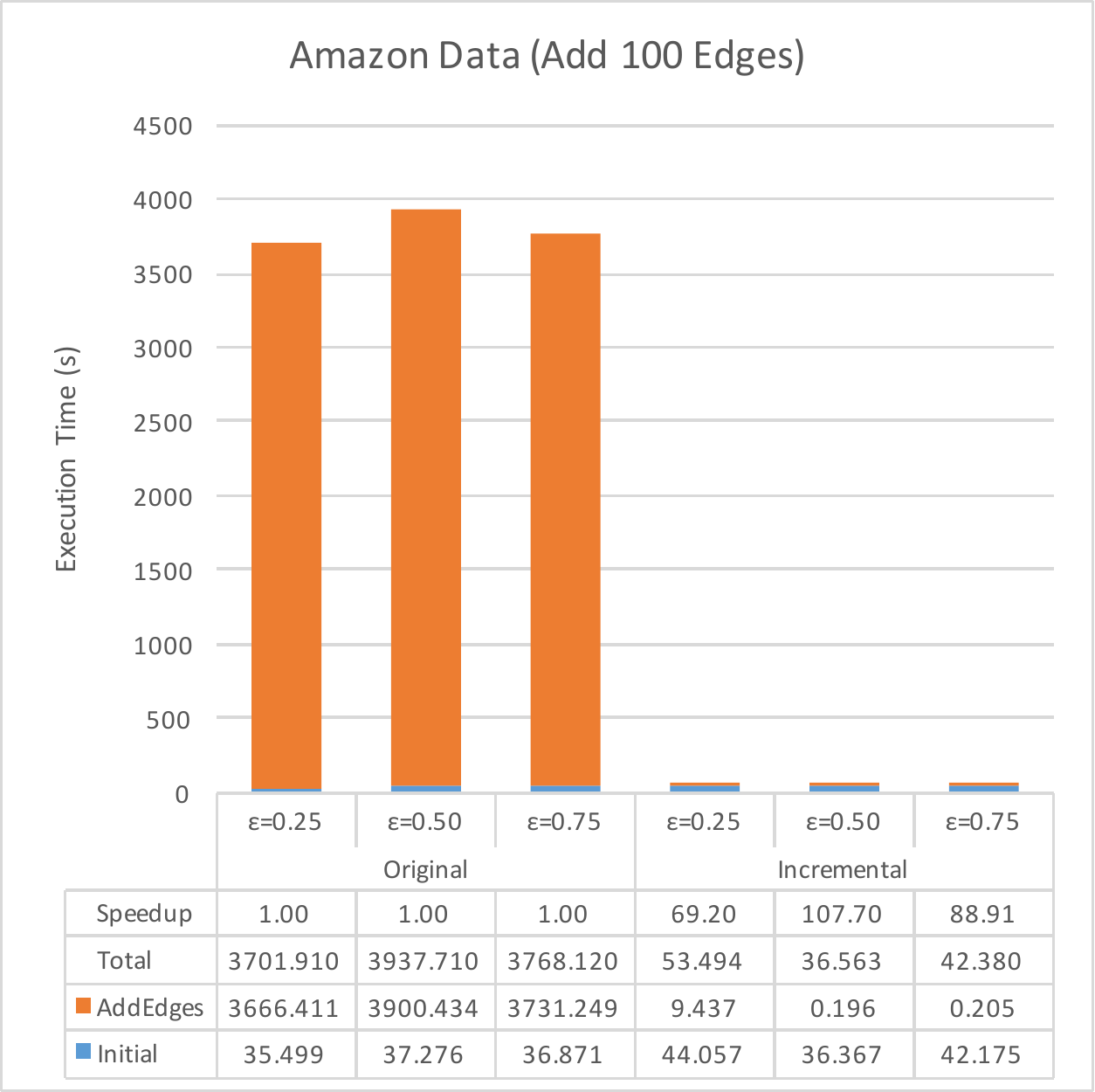}
    \caption{Elapsed time using Amazon graph data with additional 100 edges.}
    \label{fig:amazon}
\end{figure}

The execution times for community detections with additional 100 edges with Congress, IMDb and Amazon data sets are shown in Figure \ref{fig:congress}, Figure \ref{fig:imdb} and Figure \ref{fig:amazon} respectively. These figures describe the elapsed time for the first step (the lower part of each bar) to generate and analyze baseline graph objects and the changed graph with additional new 100 edges to update and analyze them (the upper part of each bar).

With Congress data set, the original DEMON algorithm took about 1.3 seconds in the first step and constantly more than one seconds in each step. The total execution time was about 130.4 seconds with the parameter $\epsilon$ used in {\it Merge} function was 0.25. In our incremental DEMON algorithm, it took about 1.29 seconds in the first step, almost same as the original algorithm. However, it took only 0.049 seconds for incremental steps to add 100 edges, resulted 1.34 seconds in total and 98.4 times faster than that of the original method.

In IMDb data set, the original algorithm took more time, about 4.79 seconds per each step, and the total execution time was more than 6 minutes. In our incremental DEMON algorithm, the first step took 4.868 seconds, but the incremental phase took only 0.96 seconds in the first step. The total execution time was only 5.828 seconds which achieved about 83.0 times faster with the parameter $\epsilon$ was 0.25. When $\epsilon$ was 0.5 or 0.75, our incremental method was around 101.5 times faster than that of the original method.

In Amazon data set, the effect of our incremental method was still remarkable.
The original DEMON algorithm almost constantly took 36.6 seconds for each step, and total execution time was more than one hour.
On the other hand, our incremental DEMON algorithm took about 44.1 seconds for the first step, but the incremental steps for additional 100 edges was only about 10.1 seconds. The total time was 54.2 seconds, 69.2 times faster than the original method.

The main reason why the incremental method took more time in the first step with Amazon data set than with Congress and IMDb data was that the number of vertices and edges were much larger, and the overheads for inner rearranging intermediate data structures such as C++ standard map structures became visible.

Even if the threshold parameter $\epsilon$ for the {\it Merge} function was changed, the total time for DEMON algorithm was not drastically changed. Generally, if $\epsilon$ is large, small communities would be frequently merged and the communities for comparison would drastically decrease. On the other hand, when $\epsilon$ is small, the number of merge processes will decrease, but many communities overlapped but not merged must be compared for every iterations. These comparison procedures could be reduced if we introduce additional data structure for comparison tables like in Figure \ref{fig:mergeOpt}. It will be our future work.

\section{Conclusion and Future Work}
We proposed an incremental community detection algorithm based DEMON algorithm for real-world network analyses. We also showed our algorithm reduced the execution time of community detection to less than single second for incrementally updated networks, and makes around 100 times faster than original method in 100 steps iterations.

Besides the speed-up of community detections for incrementally growing networks, it took much more time to generate {\it Ego Minus Ego} networks and set labels for all these networks in the first step. Moreover, it consumed much larger heap memory space than these original three functions we mentioned and it was not realistic to apply it for much larger networks with a standalone commodity machine with limited memory size.

As a future work, we will optimize {\it Ego Minus Ego} network extractions and storing functions, and improve this method on parallel and distributed-memory machines with running incremental community detections. We will only show the result on a single node, but as our next work, we will implement a distributed version of our proposed method.

In this research, we implemented and evaluated incremental method only for adding edges. We will also propose additional incremental procedures for deleting edges and vertices as another future work.

We will also need more precise evaluations about outputs like distributions of generated community sizes and precision.
Original DEMON mentioned that the distribution was better than that of other popular community detection methods.
Because our method especially {\it Label Propagation} is supposed approximation to set labels, the final result may be different from original one.


\section*{Acknowledgment}
This research was supported by JST, CREST (Research Area: Advanced Core Technologies for Big Data Integration).



\bibliographystyle{IEEEtran}

\end{document}